\documentclass[preprint,preprintnumbers,nofootinbib,aps,10pt,twocolumn]{revtex4-1}
\usepackage{amsmath,amssymb,bm,epsfig}
\usepackage{color}
\usepackage{natbib}
\usepackage{hyperref} 
\usepackage{ulem}
\usepackage{graphicx}
\usepackage{xcolor,colortbl}
\RequirePackage{lineno}
\newcommand{\nn}{\nonumber}
\newcommand{\sNN}{\sqrt{s_{\textrm{NN}}}}

\definecolor{Gray}{gray}{0.85}
\newcolumntype{a}{>{\columncolor{Gray}}c}

\def \beq{\begin{equation}}
\def \eeq{\end{equation}}
\def \beqa{\begin{eqnarray}}
\def \eeqa{\end{eqnarray}}
\def \la{\langle}
\def \ra{\rangle}
\def \l{\left(}
\def \r{\right)}
\def \l{\left(}
\def \r{\right)}

\begin{document}

\title{Freezeout systematics due to the hadron spectrum}

\author{Sandeep Chatterjee}
\email{Sandeep.Chatterjee@fis.agh.edu.pl}
\affiliation{AGH University of Science and Technology,\\ 
Faculty of Physics and Applied Computer Science,\\
al. Mickiewicza 30, 30-059 Krakow, Poland}

\author{Debadeepti Mishra}
\email{debadeepti.m@niser.ac.in}
\affiliation{School of Physical Sciences,\\ National Institute 
of Science Education and Research, HBNI,\\ Jatni-752050, India}

\author{Bedangadas Mohanty}
\email{bedanga@niser.ac.in}
\affiliation{School of Physical Sciences,\\ National Institute 
of Science Education and Research, HBNI,\\ Jatni-752050, India}

\author{Subhasis Samanta}
\email{subhasis@niser.ac.in}
\affiliation{School of Physical Sciences,\\ National Institute 
of Science Education and Research, HBNI,\\ Jatni-752050, India}

\begin{abstract}

We investigate systematics of the freezeout surface in heavy ion collisions due to the hadron 
spectrum. The role of suspected resonance states that are yet to be confirmed experimentally 
in identifying the freezeout surface has been investigated. We have studied two different 
freezeout schemes - unified freezeout scheme where all hadrons are assumed to freezeout at the 
same thermal state and a flavor dependent sequential freezeout scheme with different freezeout 
thermal states for hadrons with or without valence strange quarks. The data of mean hadron yields 
as well as scaled variance of net proton and net charge distributions have been analysed. We find 
the freezeout temperature $T$ to drop by $\sim5\%$ while the dimensionless freezeout parameters 
$\mu_B/T$ and $VT^3$ ($\mu_B$ and $V$ are the baryon chemical potential and the volume at freezeout 
respectively) are insensitive to the systematics of the input hadron spectrum. The observed hint 
of flavor hierarchy in $T$ and $VT^3$ with only confirmed resonances survives the systematics 
of the hadron spectrum. It is more prominent between $\sNN\sim10 - 100$ GeV where the maximum 
hierarchy in $T\sim10\%$ and $VT^3\sim40\%$. However, the uncertainties in the thermal parameters 
due to the systematics of the hadron spectrum and their decay properties do not allow us to make 
a quantitative estimate of the flavor hierarchy yet.

\end{abstract}

\maketitle

\section{Introduction}\label{sec.intro}
The determination of the last surface of inelastic scattering, the chemical freezeout surface (CFO), 
is an integral part of the standard model of heavy ion collision~\cite{BraunMunzinger:1995bp, Yen:1998pa, 
Cleymans:1999st, Becattini:2000jw}. An ideal gas of all the confirmed hadrons and resonances as listed by 
the Particle Data Group (PDG)~\cite{Patrignani:2016xqp} forms the Hadron Resonance Gas (HRG) model that has met 
with considerable success across a broad range of beam energies in describing the mean hadron yields~\cite{
BraunMunzinger:1995bp, Yen:1998pa, Cleymans:1999st, Becattini:2000jw, Andronic:2005yp, Chatterjee:2015fua, 
Chatterjee:2016cog} and more recently moments of conserved charges of QCD like baryon number ($B$), strangeness 
($S$) and charge ($Q$)~\cite{Karsch:2010ck, Garg:2013ata, Bhattacharyya:2013oya, Alba:2014eba} with a few 
thermal parameters. Such an analysis gives us an access to the thermodynamic state of the fireball just prior to 
freezeout. The ongoing hunt for the QCD critical point crucially depends on our knowledge of the background 
dominated by the thermal hadronic physics close to freezeout. 

The HRG partition function $Z\l T,\mu_{B,Q,S}\r$ for a thermal state at $\l T,\mu_B,\mu_Q,\mu_S\r$ 
where $T$ is the temperature and $\mu_B$, $\mu_Q$ and $\mu_S$ are the chemical potentials corresponding 
to the three conserved charges B, Q and S respectively, can be written as
\beqa
\ln Z &=& \sum_i\ln Z_i\l T,\mu_{B,Q,S}\r\label{eq.lnZ}
\eeqa
where $Z_i$ is the single particle partition function corresponding to the $i$th hadron species written as
\beqa
\ln Z_i\l T,\mu_{B,Q,S}\r &=& VT^3\frac{ag_i}{2\pi^2}\int dpp^2/T^3\nn\\
&&\times\ln\l1+ae^{-\l\sqrt{\l p^2+m_i^2\r}+\mu_i\r/T}\r\label{eq.lnZi}
\eeqa
where $a=-1(+1)$ for mesons (baryons), $g_i$ and $m_i$ refer to the degeneracy factor and mass of the $i$th 
hadron species and $\mu_i$ is its hadron chemical potential which within a complete chemical equilibrium 
scenario is written as 
\beqa
\mu_i &=& B_i\mu_B+Q_i\mu_Q+S_i\mu_S\label{eq.mui}
\eeqa
where $B_i$, $Q_i$ and $S_i$ are the baryon number, charge and strangeness of the $i$th hadron species. 

The sum in Eq.~\ref{eq.lnZ} runs over all the established resonances from the PDG. However, quark 
models~\cite{Capstick:1986bm, Ebert:2009ub} and studies on the lattice~\cite{Edwards:2012fx} predict many more 
resonances than that have been confirmed so far in experiments. It has been pointed out in studies based on 
comparison between QCD thermodynamics on the lattice and HRG that these resonances could have significant 
contribution to several thermodynamic quantities~\cite{Bazavov:2014xya, Alba:2017mqu} and influence the extraction 
of the freezeout parameters within the HRG framework~\cite{Bazavov:2014xya}. There have been interesting 
studies on the status and influence of the systematics of the hadron spectrum on several quanitites~\cite{
Broniowski:2000bj,Broniowski:2004yh,NoronhaHostler:2007jf,Chatterjee:2009km,Lo:2015cca,RuizArriola:2016qpb,
Noronha-Hostler:2016ghw,Begun:2017yus}. In this work, we have studied the systematics in the determination of 
the freezeout surface within the HRG framework due to the uncertainties over the hadron spectrum.

\section{Extracting Chemical Freezeout Surface}
\label{sec.cfo}
The standard pratice has been to extract the CFO parameters by fitting the mean hadron yields. The primary 
yields $N^p_i$ are obtained from Eq.~\ref{eq.lnZ} as follows
\beqa
N^p_i &=& \frac{\partial}{\partial\mu_i}\ln Z\label{eq.npi}
\eeqa
while the total yields of the stable hadrons that are fitted to data are obtained after adding the secondary 
contribution from the resonance decays to their primary yields
\beqa
N^t_i &=& N^p_i + \sum_j BR_{j\rightarrow i}N^p_j\label{eq.nti}
\eeqa
where $BR_{j\rightarrow i}$ refers to the branching ratio (BR) of the decay of the $j$th to $i$th hadron species.

In this work we characterise the freezeout surface by three parameters of which only $T$ has dimension. The 
other two are suitably scaled dimensionless paramteters $\mu_B/T$ and $VT^3$. While $\mu_B/T$ controls the 
baryon fugacity factor, $VT^3$ can be interpreted as the effective phase space volume occupied by the HRG at 
freezeout. The masses of the hadrons are the relevant scales in this problem. They decide the freezeout $T$. 
Thus, it is natural that systematic variation of the hadron spectrum will result in corresponding variation 
of the freezeout $T$. On the other hand, the influence of the systematics of the hadron spectrum on the 
dimensionless parameters $\mu_B/T$ and $VT^3$ is expected to be lesser. This motivates us to work with $\l T, 
\mu_B/T, VT^3\r$ instead of the standard choice of $\l T, \mu_B, V\r$. The other parameter that is often used 
in literature, the strangeness understauration factor $\gamma_S$ has been taken to be unity here. $\mu_S$ and 
$\mu_Q$ are solved consistently from the strangeness neutrality condition and demanding the ratio of net $B$ to 
$Q$ be equal to that of the colliding nuclei (this ratio $\sim2.5$ for Au and Pb nuclei which we consider here)
\beqa
\text{Net }S &=& 0\label{eq.netS}\\
\text{Net }B/\text{Net }Q &=& 2.5\label{eq.muQ}
\eeqa
As is evident from Eq.~\ref{eq.nti}, knowledge of the BRs is essential to compute the contribution of 
the secondary yield which is the feeddown from the heavier unstable resonances to the observed hadrons. As a result, 
extraction of the freezeout surface based on the hadron yield data suffer from systematic uncertainties 
of the decay properties of these addtional resonances.

The freezeout surface can also be estimated by comparing higher moments of the conserved charges in 
experiment and theory~\cite{Karsch:2010ck, Gavai:2010zn, Bazavov:2012vg, Borsanyi:2013hza, Garg:2013ata, Bhattacharyya:2013oya, 
Alba:2014eba, Bazavov:2015zja}. One of the important advantage in using the fluctuations of conserved charges over 
hadron yields in estimating the freezeout surface is that it is enough to know only the quantum numbers of these 
unconfirmed states. Decays under strong interactions should conserve the charges $B$, $Q$ and $S$. Thus the conserved 
charge susceptibilities are not influenced by the systematic uncertainties of the BRs of the unconfirmed resonances. 

On the theoretical side, it is straightforward to compute the conserved charge susceptibilities $\chi^{ijk}_{BQS}$ 
of order $\l i+j+k\r$ from the partition function
\beqa
\chi^{ijk}_{BQS} &=& \frac{\partial^{i+j+k}\l P/T^4\r}
{\partial^i\l\mu_B/T\r\partial^j\l\mu_Q/T\r\partial^k\l\mu_S/T\r}\label{eq.sus}
\eeqa
where the pressure $P$ is obtained from
\beqa
P &=& \frac{T}{V}\ln Z\label{eq.p}
\eeqa
The above susceptibilities computed in a model can then be converted easily to moments for a comparison 
with the measured data. For example, the mean $M$ and variance $\sigma^2$ of the conserved charge distribution 
has one-to-one correspondence with the first two orders of susceptibility of the respective charge $c$
\beqa
M_c &=& \la N_c\ra = VT^3\chi^1_c\label{eq.mean}\\
\sigma^2_c &=& \la\l N_c-\la N_c\ra\r^2\ra = VT^3\chi^2_c\label{eq.variance}
\eeqa
with $N_c$ being the observed net charge of type $c$ in an event while $\la N_c\ra$ is the ensemble average.  
We have evaluated the susceptibilities within HRG and estimated the influence of the missing resonances to 
the extraction of freezeout parameters thereof. It has been found that scaled variance $\sigma^2/M$ of net $Q$ 
and net $B$ are well described within the HRG framework while higher moments like skewness and kurtosis show 
up discrepancises, particularly at lower energies~\cite{Gupta:2011wh,Alba:2014eba,Adak:2016jtk}. These higher 
moments are also sensitive to non-ideal corrections like incorporating repulsive and attractive interactions 
within the HRG framework~\cite{Albright:2015uua,Vovchenko:2016rkn,Huovinen:2017ogf,Vovchenko:2017xad}. Hence, 
in this study we stick to $\sigma^2/M$ of net $Q$ and $B$ to ascertain the influence of the systematics of 
the hadron spectrum on the freezeout surface extracted from the data on conserved charge fluctuation. 

On the experimental front, several uncertainties can creep into the measurement of conserved charge fluctuations. 
The acceptance cuts in transverse momentum and rapidity is one of them~\cite{Garg:2013ata}. Also, the neutral particles are 
not detected which means net-proton fluctuations only act as an approximate proxy for net $B$~\cite{Kitazawa:2012at}. 
Currently, there is a tension between the PHENIX~\cite{Adare:2015aqk} and STAR~\cite{Adamczyk:2013dal,Adamczyk:2014fia} 
measurements for net charge fluctuation. In this work, we have extracted the freezeout parameters for STAR data alone. 
We expect the dependence of the freezeout parameters on the uncertainties of the hadron spectrum to be similar for 
STAR and PHENIX data.

While a single unified freezeout picture (1CFO) provides a good qualitative description across a broad range of beam 
energies, recent studies have shown that a natural step beyond 1CFO would be to consider flavor hierarchy in 
freezeout (2CFO), based on various arguments like flavor hierarchy in QCD thermodynamic quantities on the 
lattice~\cite{Bellwied:2013cta}, hadron-hadron cross sections~\cite{Chatterjee:2013yga, Bugaev:2013sfa} and melting 
of in-medium hadron masses~\cite{Torres-Rincon:2015rma}. We have analysed the yield data in both the freezeout 
schemes: 1CFO and 2CFO. However, for the conserved charge fluctuation study, currently only data for moment of net 
proton (proxy for net baryon)~\cite{Adamczyk:2013dal} and net charge~\cite{Adamczyk:2014fia} are available. The net 
$B$ and net $Q$ fluctuations are dominated by the non-strange sector as the lightest hadrons contributing to these 
quantities are non-strange. Thus, the analysis of the data on the conserved charge fluctuations is not sensitive to 
the thermal state of the strange sector. Hence, we have analysed the data on higher moments of the conserved charges 
only within 1CFO.

\section{Hadron spectrum}\label{sec.spectrum}

\begin{table} [ht]
\begin{center}
\begin{tabular}{|c|c||c|c|} 
\hline
\multicolumn{2}{|c||}{Mesons} & \multicolumn{2}{c|}{Baryons}\\\hline
$h_1\l1380\r$ & $f_2\l1430\r$ & $N\l1860\r$ & $N\l1880\r$\\\hline
$f_1\l1510\r$ & $f_2\l1565\r$ & $N\l1895\r$ & $N\l1895\r$\\\hline
$\rho\l1570\r$& $h_1\l1595\r$ & $N\l2000\r$ & $N\l2040\r$\\\hline
$a_1\l1640\r$ & $f_2\l1640\r$ & $N\l2060\r$ & $N\l2100\r$\\\hline
$a_2\l1700\r$ & $\eta\l1760\r$& $N\l2120\r$ & $N\l2300\r$\\\hline
$f_2\l1810\r$ & $a_1\l1420\r$ & $N\l2570\r$ & $N\l2700\r$\\\hline
$\eta_2\l1870\r$ & $\rho\l1900\r$ & $\Delta\l1750\r$ & $\Delta\l1900\r$\\\hline
$f_2\l1910\r$ & $a_0\l1950\r$ & $\Delta\l1940\r$ & $\Delta\l2000\r$ \\\hline
$\rho_3\l1990\r$ & $f_0\l2020\r$ & $\Delta\l2150\r$ & $\Delta\l2200\r$ \\\hline
$\pi_2\l2100\r$ & $f_0\l2100\r$ & $\Delta\l2300\r$ & $\Delta\l2350\r$ \\\hline
$f_2\l2150\r$ & $\rho\l2150\r$ & $\Delta\l2390\r$ & $\Delta\l2400\r$ \\\hline
$f_0\l2200\r$ & $f_4\l2220\r$ & $\Delta\l2750\r$ & $\Delta\l2950\r$\\\hline
$\eta\l2225\r$ & $\rho_3\l2250\r$ &  & \\\hline
$f_4\l2300\r$ & $f_0\l2330\r$ & & \\\hline
$\rho_5\l2350\r$ & $a_6\l2450\r$ & & \\\hline
$f_6\l2510\r$ &  & & \\\hline\hline
$K\l1460\r$ & $K_2\l1580\r$ & $\Lambda\l1710\r$ & $\Lambda\l2000\r$\\\hline
$K\l1630\r$ & $K_1\l1650\r$ & $\Lambda\l2020\r$ & $\Lambda\l2050\r$\\\hline
$K\l1830\r$ & $K^*_0\l1950\r$ & $\Lambda\l2325\r$ & $\Lambda\l2585\r$ \\\hline
$K^*_2\l1980\r$ & $K_2\l2250\r$ & $\Sigma\l1480\r$ & $\Sigma\l1560\r$ \\\hline
$K_3\l2320\r$ & $K^*_5\l2380\r$ & $\Sigma\l1580\r$ & $\Sigma\l1620\r$ \\\hline
$K_4\l2500\r$ & $K\l3100\r$ & $\Sigma\l1690\r$ & $\Sigma\l1730\r$\\\hline
 &  & $\Sigma\l1770\r$ & $\Sigma\l1840\r$\\\hline
 &  & $\Sigma\l1880\r$ & $\Sigma\l1900\r$\\\hline
 &  & $\Sigma\l1940\r$ & $\Sigma\l2000\r$\\\hline
 &  & $\Sigma\l2070\r$ & $\Sigma\l2080\r$\\\hline
 &  & $\Sigma\l2100\r$ & $\Sigma\l2455\r$\\\hline
 &  & $\Sigma\l2620\r$ & $\Sigma\l3000\r$\\\hline
 &  & $\Sigma\l3170\r$ & $\Xi\l1620\r$\\\hline
 &  & $\Xi\l2120\r$ & $\Xi\l2250\r$\\\hline
 &  & $\Xi\l2370\r$ & $\Xi\l2500\r$\\\hline
 &  & $\Omega\l2380\r$ & $\Omega\l2470\r$\\\hline
\end{tabular}
\caption{List of additional resonances in PDG-2016+ that are not included in PDG-2016. This consists of 
1- and 2- stars status baryons and unmarked mesons from PDG 2016~\cite{Patrignani:2016xqp} that are yet 
to be confirmed experimentally.}
\end{center}
\label{tab.2016+}
\end{table}

We have performed our analysis with two different hadron spectrum: the first set consists of only the confirmed 
hadrons and resonances from PDG 2016~\cite{Patrignani:2016xqp}. This includes all the mesons listed in the Meson 
Summary Table~\cite{Patrignani:2016xqp} that are marked confirmed and all baryons in the Baryon Summary 
Table~\cite{Patrignani:2016xqp} with 3- or 4- star status that are considered established. This we refer to as PDG-2016. 
The second set includes all the resonances from PDG-2016 as well as the other unmarked mesons from the Meson Summary 
Table and baryons from the Baryon Summary Table with 1- or 2- star status that await confirmation. This set is referred 
to as PDG-2016+. We have listed all the additional resonances in PDG-2016+ in Table~\ref{tab.2016+}. We consider 
resonances with only up, down and strange flavor valence quarks. It has been shown that the PDG-2016+ provides a 
satisfactory description in the hadronic phase of continuum lattice estimates of most thermodynamic 
quantities~\cite{Alba:2017mqu}. 

As emphasised earlier, BRs for unstable resonances are an important ingredient to compute the total hadron yields. 
While the PDG provides the BRs for most of the confirmed resonances, those for the 
unconfirmed resonances are missing. Hence, we have to make an assumption of their decay properties. The resonances of 
the same family have similar decay properties. Here, we assign to an unknown resonance $R$ the decay properties of the 
known resonances with same quantum numbers as $R$ and immediate to $R$ in mass. For a systematic dependence on these 
unknown BRs, we perform our analysis with BRs taken from different resonances that are lighter as well as heavier to 
$R$. This systematic variation results in large variation in $\chi^2$ and larger error bars in the fits with the 
PDG-2016+ spectrum as compared to PDG-2016. This also restricts us from including further unknown states predicted 
by theoretical studies as the systematic uncertainties will be too large to draw any physics conclusions.

\begin{figure}
 \begin{center}
 \includegraphics[scale=0.42]{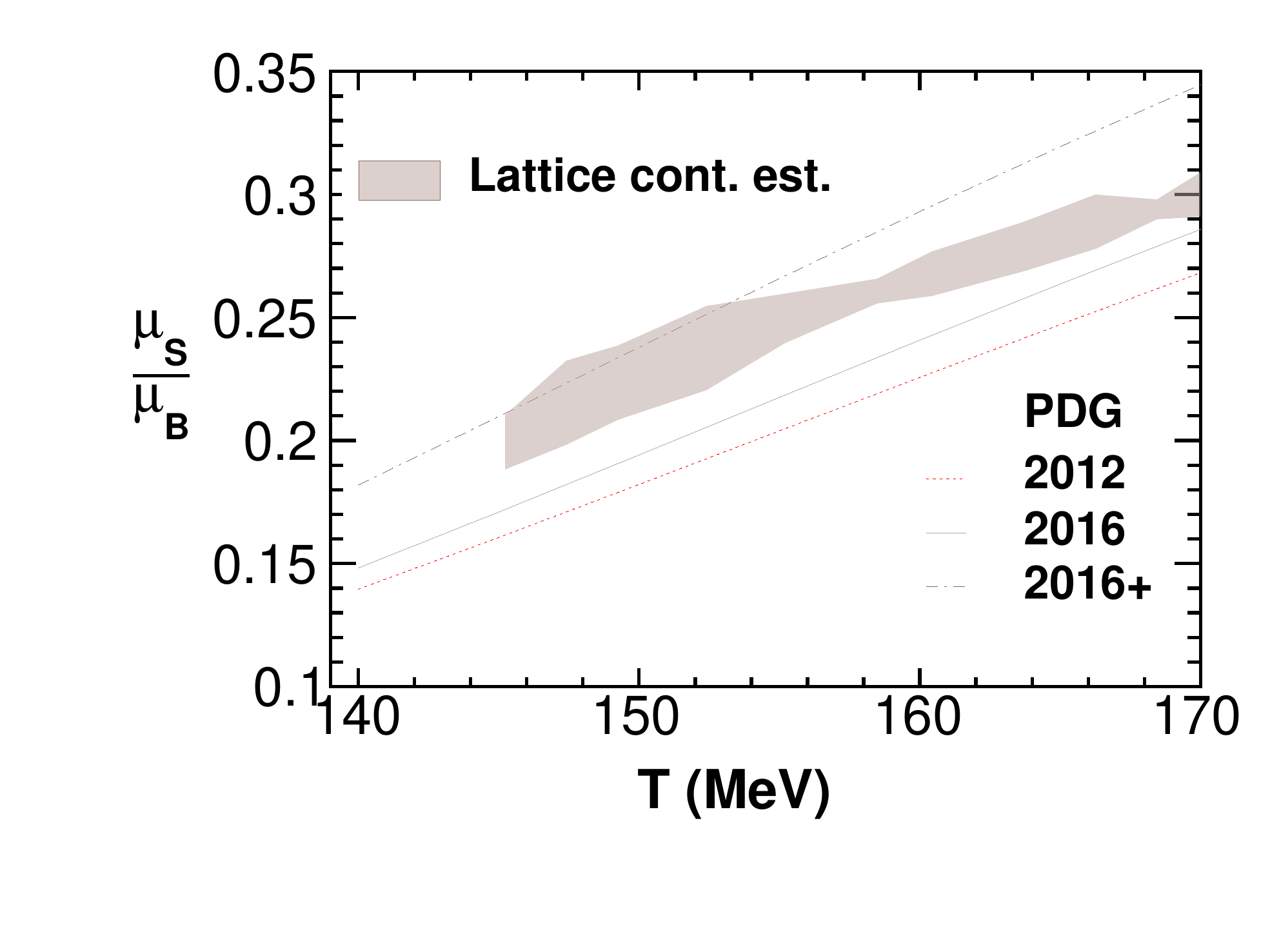}
 \caption{(Color online) Leading order $\mu_S/\mu_B$ from continuum estimate of lattice~\cite{Bazavov:2014xya}
 compared to that of HRG with hadron spectrum from PDG 2012, 2016 and 2016+.}
 \label{fig.musbyb}
 \end{center}
\end{figure}

The $\mu_S$ extracted from the strangeness neutrality condition in Eq.~\ref{eq.netS} is sensitive to the strange 
hadron spectrum~\cite{Bazavov:2014xya}. There has been studies on the lattice of the ratio $\mu_S/\mu_B$ that 
constrains the hadron spectrum, specifically the strange sector~\cite{Bazavov:2014xya, Alba:2017mqu}. It has been 
shown that the PDG 2012 hadron spectrum underestimates the ratio $\mu_S/\mu_B$ at a given $T$. This can be addressed 
by including more states that are yet to be confirmed experimentally but predicted by quark model based studies or 
on the lattice. The PDG 2016 has confirmed several new resonances while having others in the list of unconfirmed. 
We have analysed their influence on this ratio by computing the same in PDG-2012, PDG-2016 and PDG-2016+. The ratio 
rises as we go from PDG-2012 to 2016 to 2016+, underlining the significance of these new resonances. This is particularly 
important to address the important issue of flavor hierarchy. As argued in Ref.~\cite{Bazavov:2014xya}, the same value 
of $\mu_S/\mu_B$ is realised at a lower $T$ on the lattice as well as in HRG with additional resonances apart from those 
listed in the 2012 version of PDG (PDG-2012)~\cite{Beringer:1900zz}. This brings down the strange freezeout $T$ that can 
possibly modify the flavor hierarchy seen in the earlier 2CFO HRG fit with the PDG-2012~\cite{Chatterjee:2013yga}. We 
see in Fig.~\ref{fig.musbyb} that the results for the PDG-2016 and 2016+ flank the continuum estimate of lattice. Thus, 
the PDG-2016+ cures the incorrect estimate of $\mu_S$ by PDG-2012 that lowers the freezeout $T$ extracted for the strange 
sector. This makes it interesting to check the status of flavor hierarchy with PDG-2016+.

\section{Results}\label{sec.result}

\begin{figure*}[t]
 \begin{center}
 \includegraphics[scale=0.4]{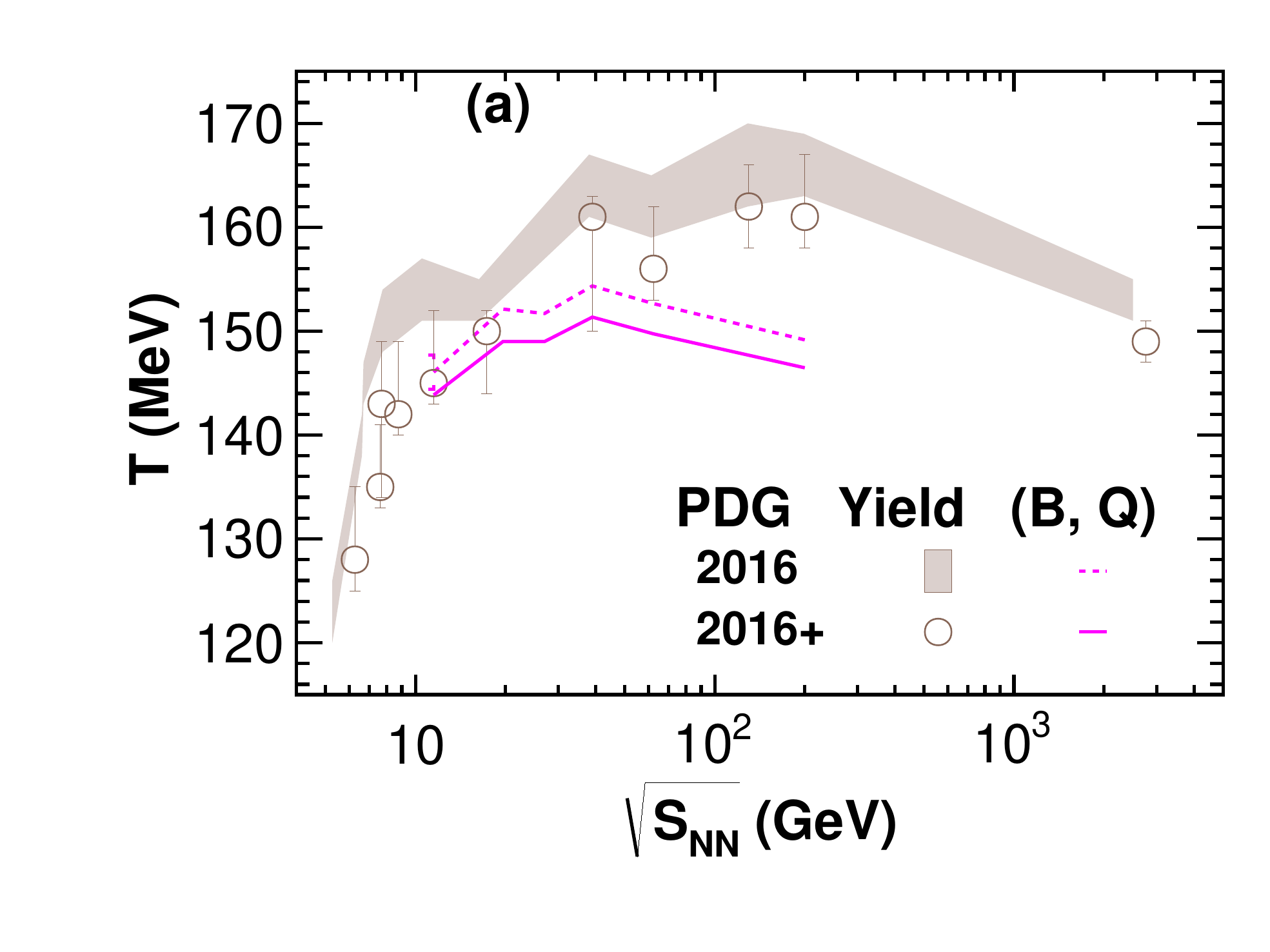}
 \includegraphics[scale=0.4]{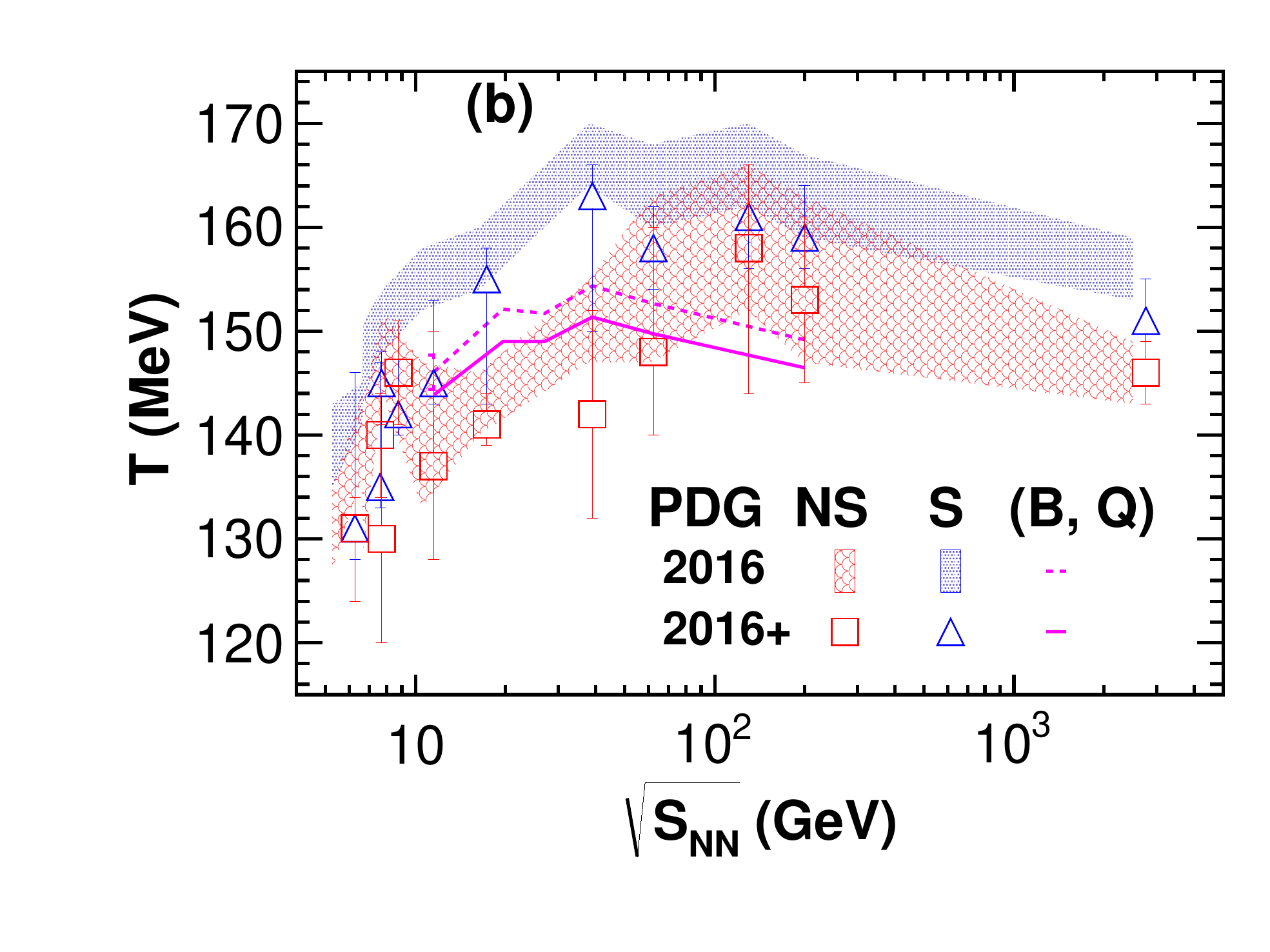}\\
 \includegraphics[scale=0.4]{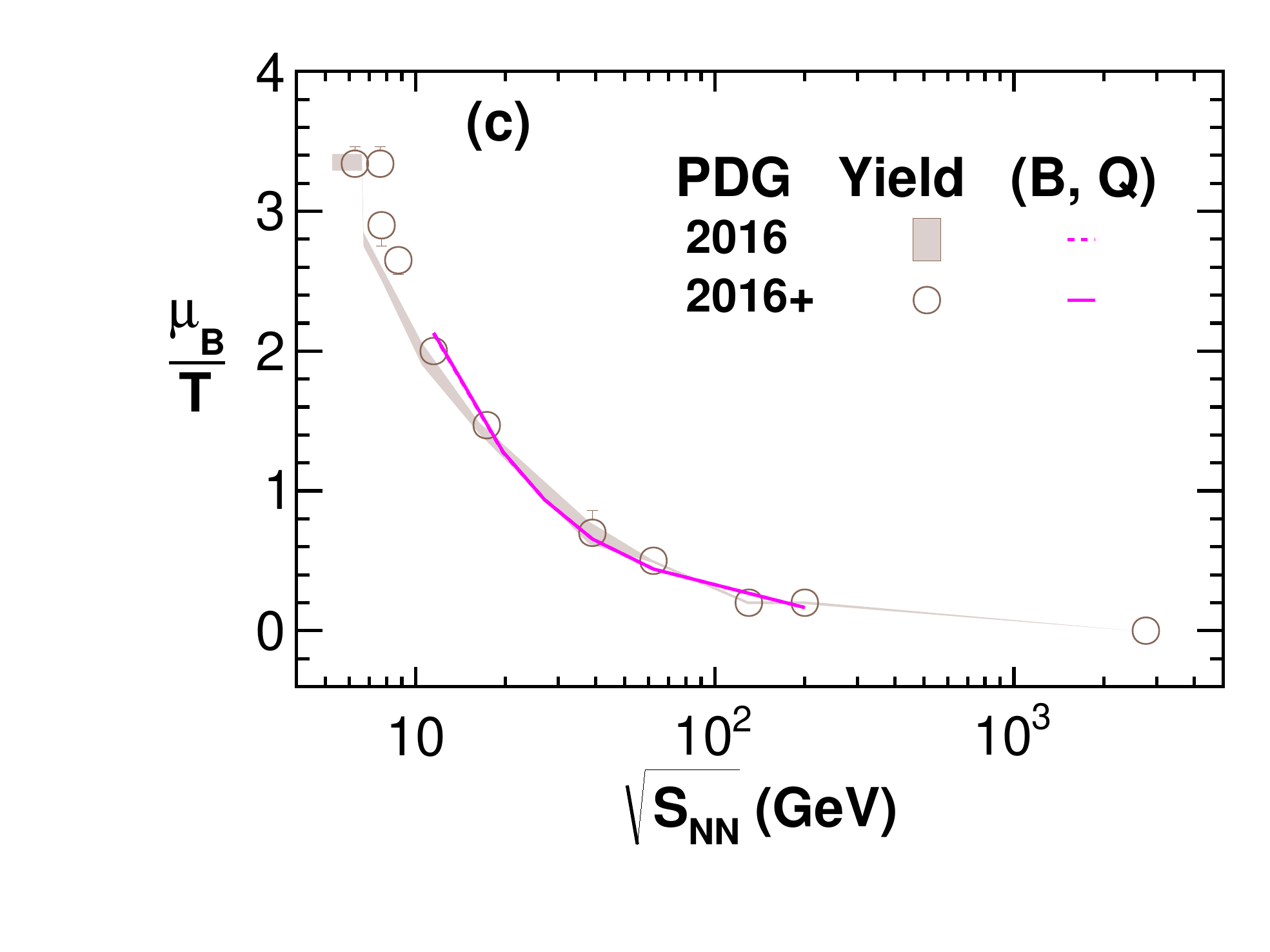}
 \includegraphics[scale=0.4]{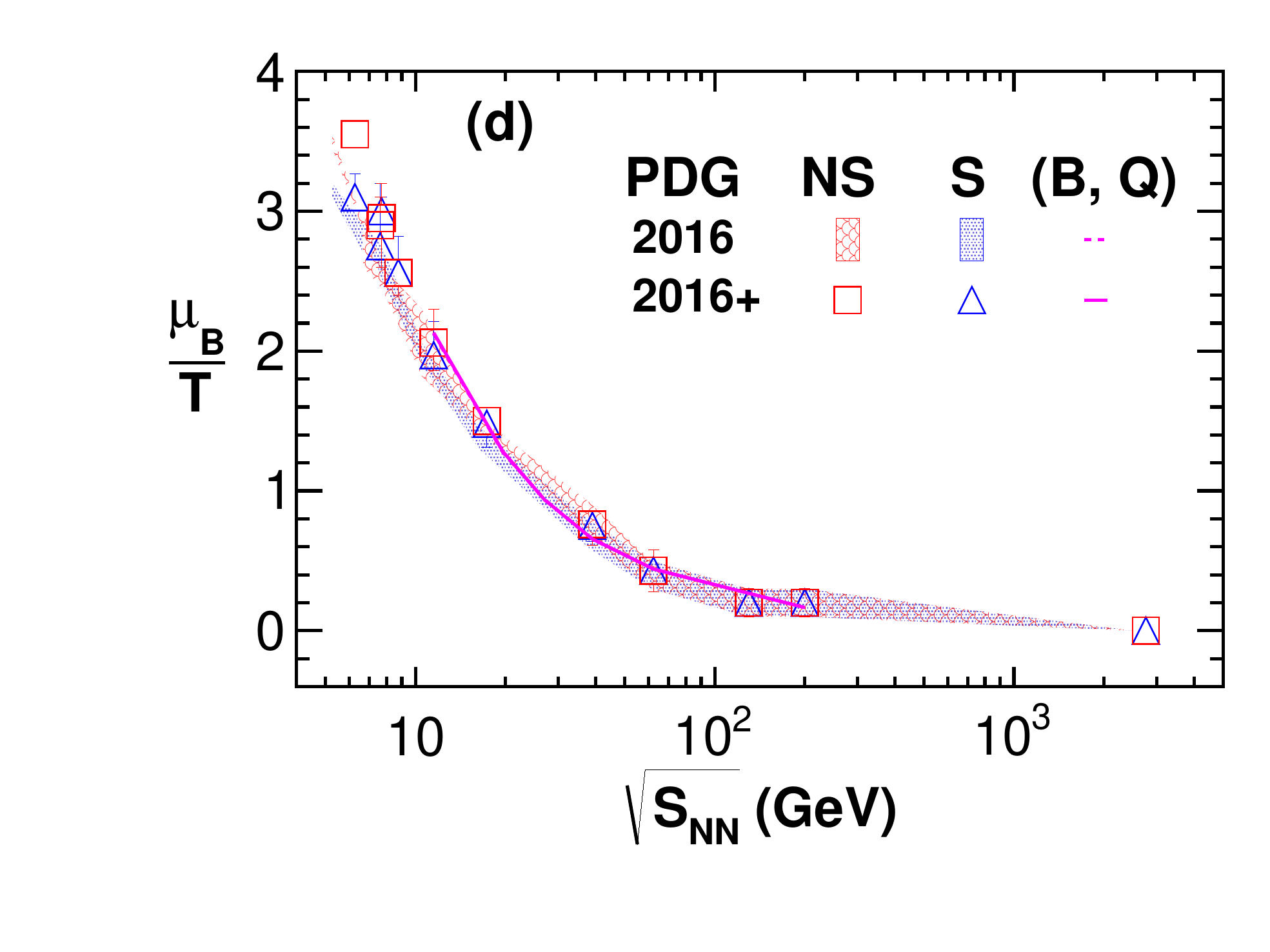}\\
 \includegraphics[scale=0.4]{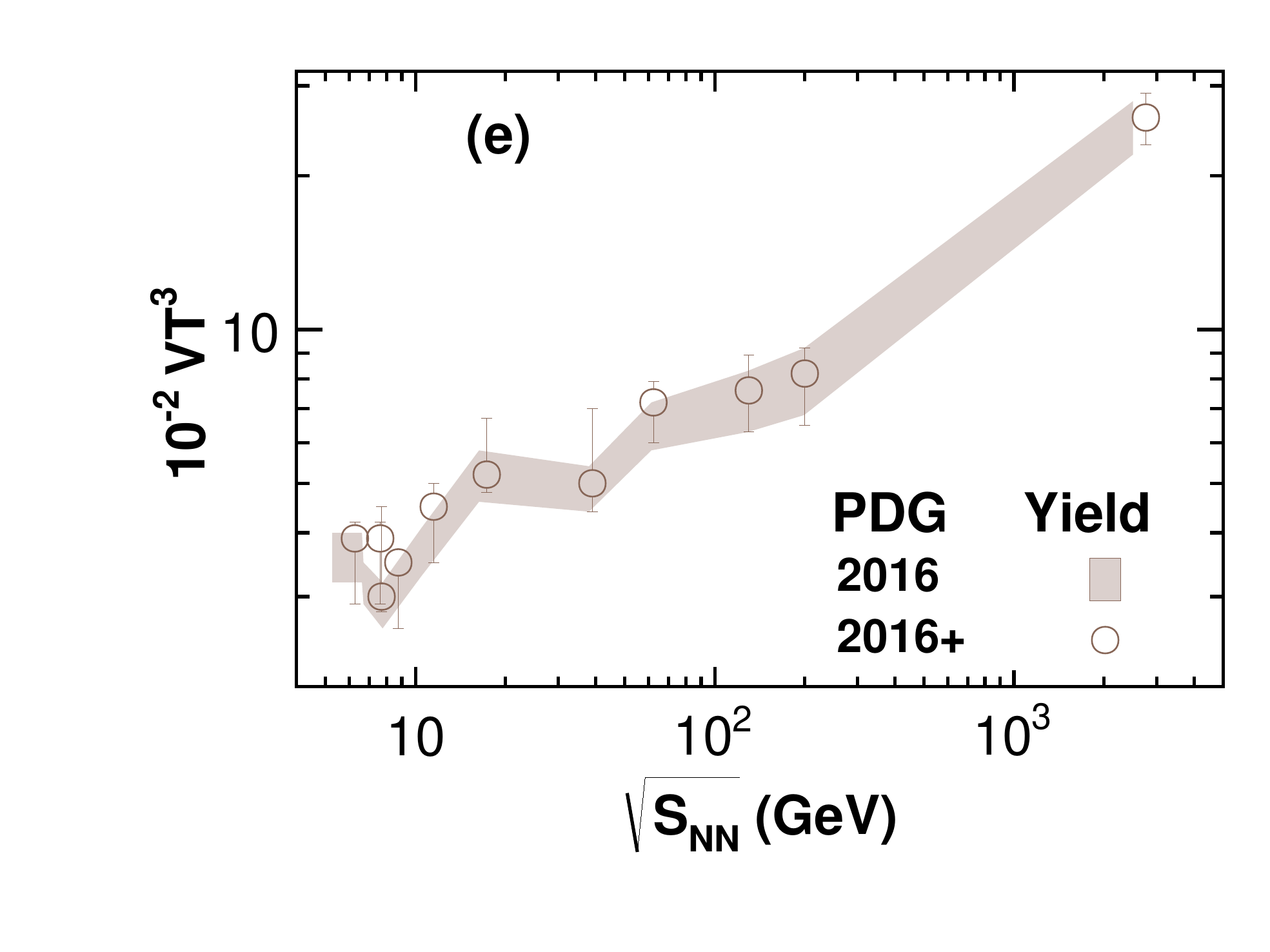}
 \includegraphics[scale=0.4]{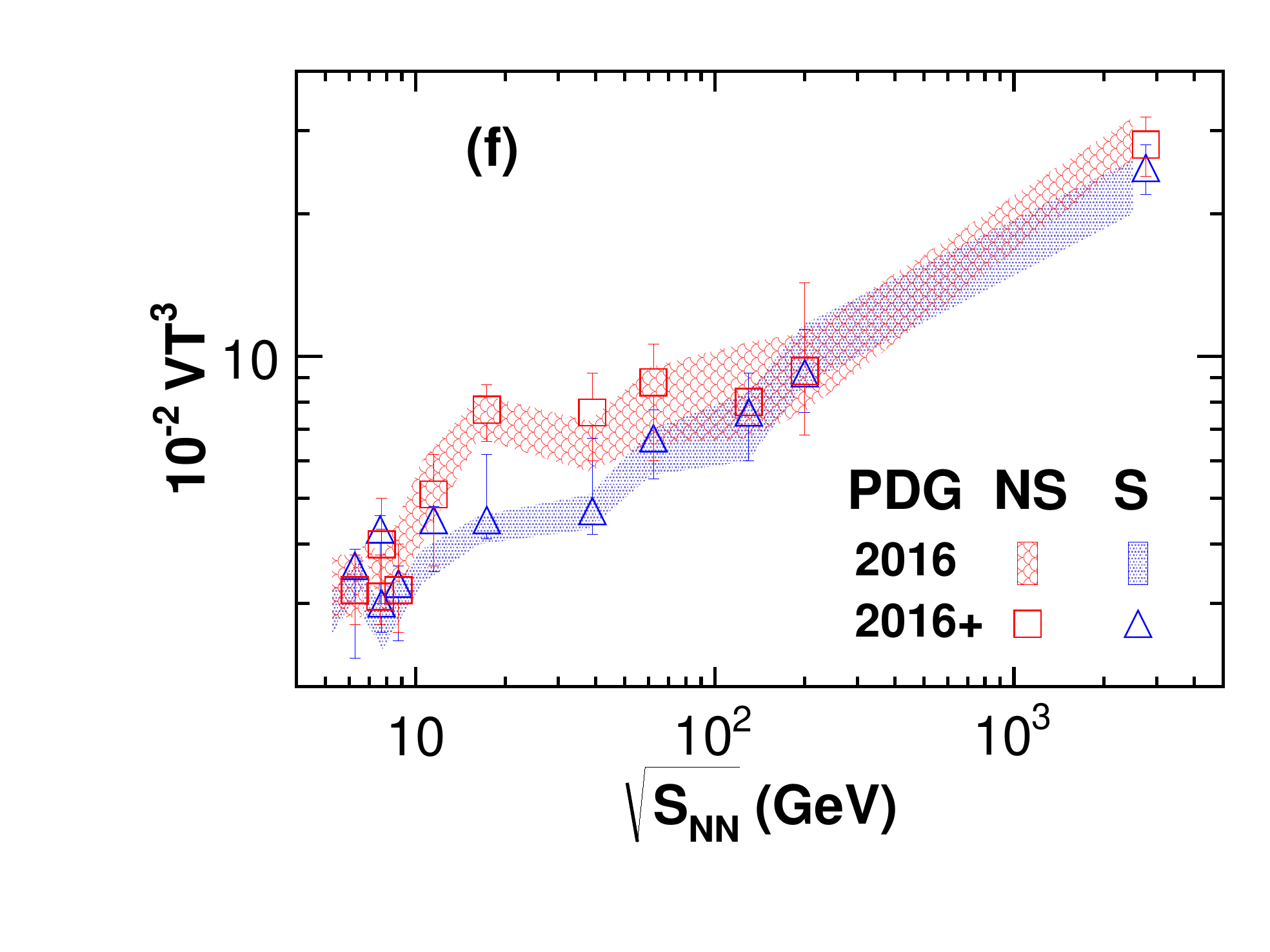}
 \caption{(Color online) The extracted thermal parameters in 1CFO (left) and 2CFO (right) schemes with PDG-2016 
 and PDG-2016+ hadron spectra. $\l B, Q\r$ refer to the use of data on conserved charges net $B$ and net $Q$ to extract 
 freezeout parameters.}
 \label{fig.cfo}
 \end{center}
\end{figure*}

We have extracted the freezeout parameters from the data on mean mid-rapidity hadron 
yields~\cite{Alt:2007aa,Alt:2005gr,Alt:2008qm,Alt:2008iv,Afanasiev:2002mx,Alt:2006dk,Alt:2004kq,
Das:2012yq,Abelev:2008ab,Abelev:2008aa,Aggarwal:2010ig,Adler:2002xv,Adams:2003fy,Adcox:2003nr,Adcox:2002au,
Adams:2004ux,Adler:2003cb,Adams:2006ke,Abelev:2013vea, Abelev:2014uua,Abelev:2013xaa, ABELEV:2013zaa} 
as well as scaled variance of net $B$ and net $Q$~\cite{Adamczyk:2013dal,Adamczyk:2014fia}. The 
thermal parameters extracted from the 1CFO (left) and 2CFO (right) schemes have been shown in Fig.~\ref{fig.cfo}. 
In the top panels, the extracted $T$ has been plotted for different beam energies. We find 
that in 1CFO as well as non-strange and strange freezeout temperatures in 2CFO, the extracted 
temperatures seem to lower as we include more resonances. However, the systematic 
uncertainties being large do not allow us to make a conclusive statement. Hence, we 
have also performed the fits to the data of conserved charge fluctuation which is shown in dashed 
and solid lines in pink for PDG-2016 and PDG-2016+ respectively. The $T$ extracted from the 
conserved charge fluctuation data clearly reveal signature of cooling on addition of more resonances.
As has been observed earlier~\cite{Alba:2014eba}, the $T$ obtained from fit to data on higher moments 
seem to be lower than that obtained from yields. The tension between these tempertures is bit lessened 
for $\sNN<20$ GeV. In 2CFO, the $T$ extracted from fluctuation data for $B$ and $Q$ (which is dominated 
by the non-strange sector as both the lightest charged hadron, $\pi$ and baryon, $p$ are non-strange 
hadrons) is clearly closer to the non-strange $T$ for PDG-2016 as well as PDG-2016+. 

In the middle panels we have plotted freezeout $\mu_B/T$ at different $\sNN$. This 
baryon fugacity parameter seems to be quite stable across freezeout schemes, experimental data of 
yields and conserved charge fluctuation as well as the different hadron spectrum. For $\sNN<10$ GeV, 
in a highly baryonic fireball, there seems to be a mild dependence on flavor as well as the hadron 
spectrum. Finally, the phase space volume factor $VT^3$ where $V$ is the coordinate space freezeout
volume has been plotted in the bottom panels. This paramter also seems quite stable against addition 
of extra resonances in PDG-2016+. However, unlike $\mu_B/T$ this has a clear flavor hierarchy 
structure similar to $T$. The non-strange phase space volume comes out to be larger than that of the 
strange phase space volume mostly.

\begin{figure*}
 \begin{center}
 \includegraphics[scale=0.42]{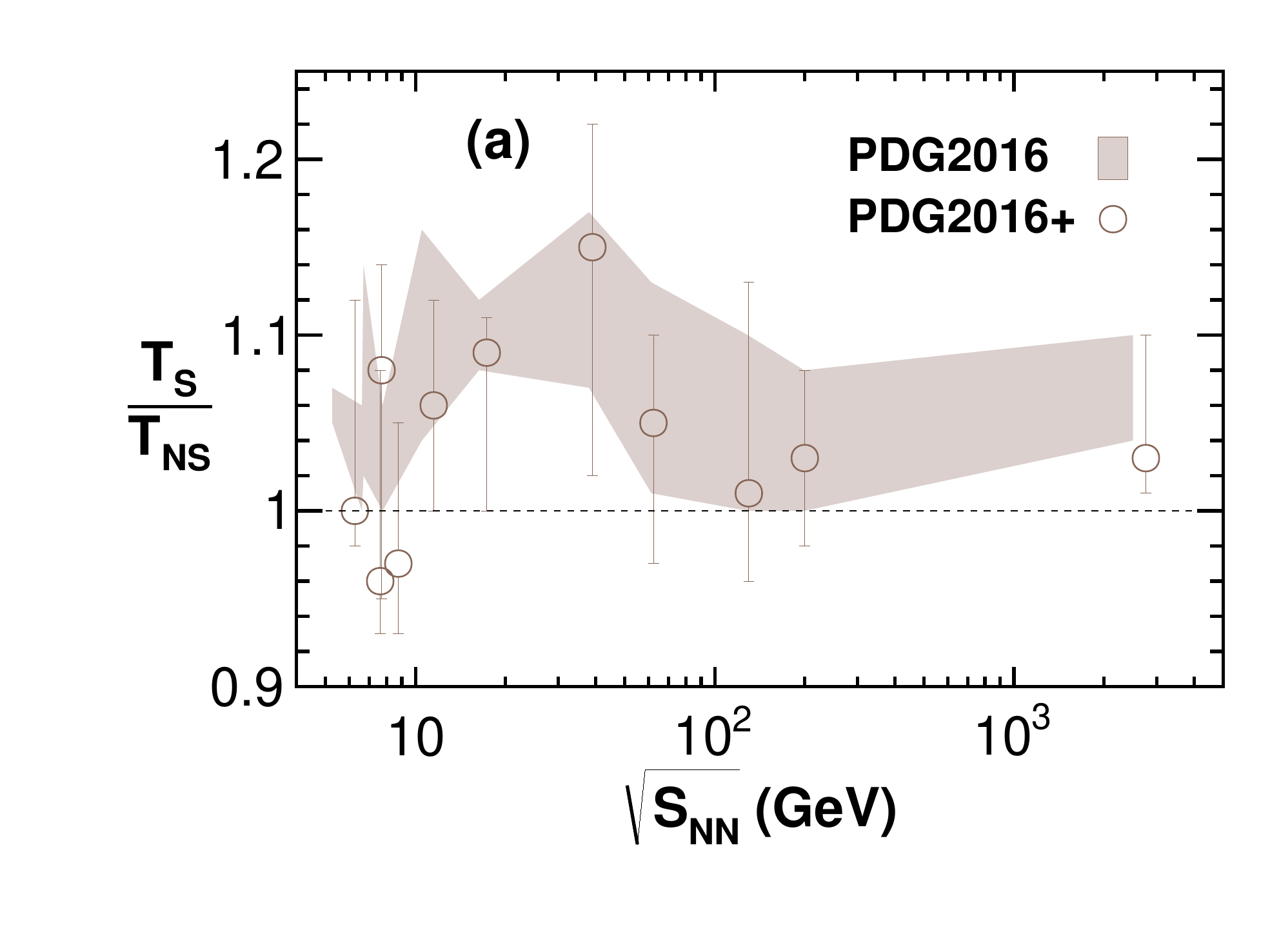}
 \includegraphics[scale=0.42]{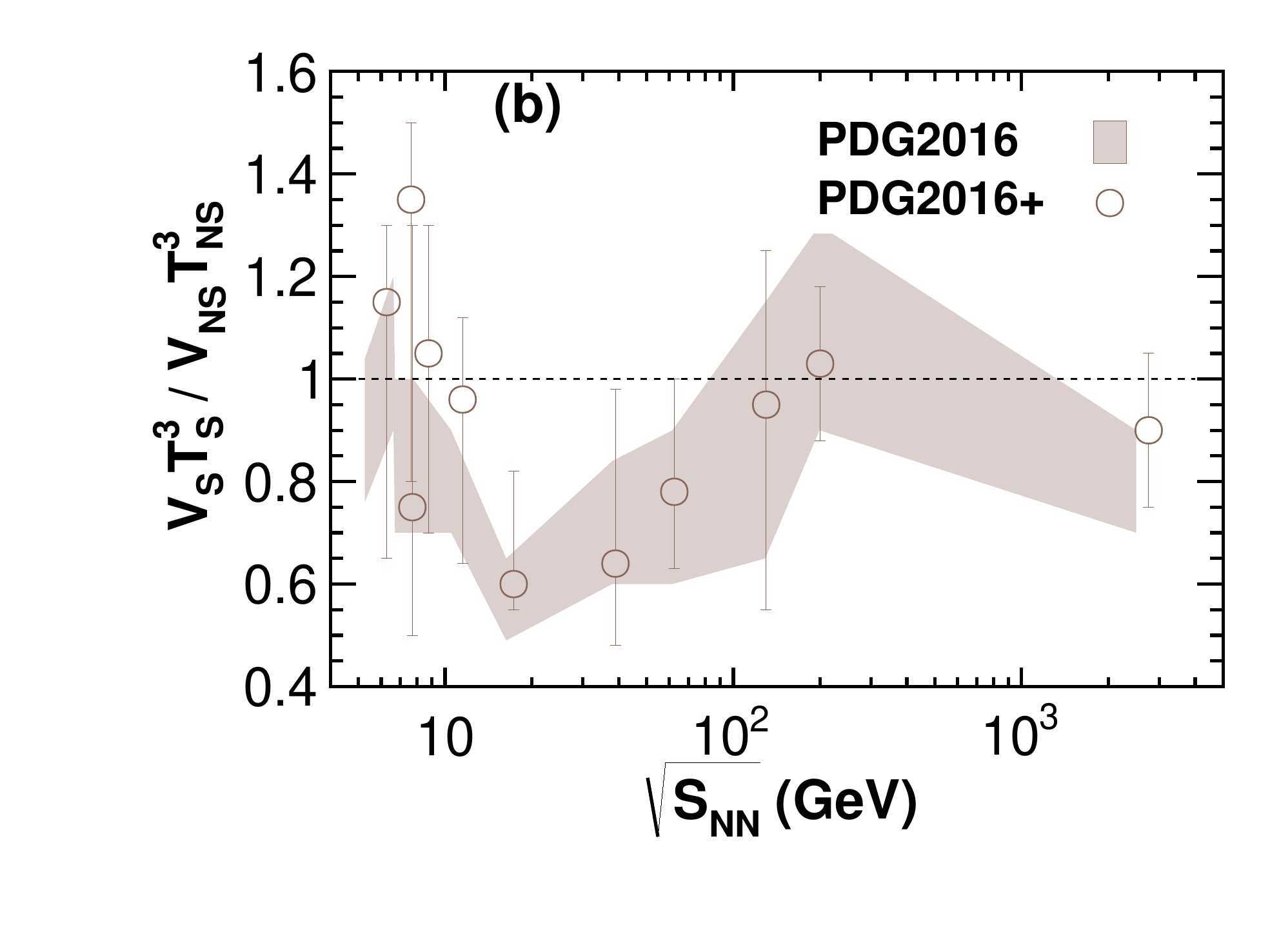}
 \caption{(Color online) Flavor hierarchy in freezeout $T$ (left) and $VT^3$ (right) and its dependence 
 on the hadron spectrum.}
 \label{fig.sbyns}
 \end{center}
\end{figure*}

There are correlations between the strange and non-strange freezeout parameters. Hence, it might be 
misleading to comment on the flavor hierarchy from Fig.~\ref{fig.cfo}. Thus, we have extracted the 
strange to non-strange freezeout parameters directly from the fits as well. These have been plotted 
in Fig.~\ref{fig.sbyns}. We find that the flavor hierarchy 
in $T$ obtained with PDG-2016 is same as that obtained earlier with PDG-2012~\cite{Chatterjee:2013yga}. 
Further, for PDG-2016+, the central values clearly indicate the flavor hierarchy. The large 
error bars (due to the uncertainties coming from the systematic variation of the decay properties of the 
unconfirmed resonances) do not allow us to make a quantitative estimate of this 
hierarchy. However, even afer including the error bars, qualitatively it is possible to state 
that data favors $T_s/T_{ns}>1$ while $V_sT_s^3/V_{ns}T_{ns}^3<1$ across most beam energies. At LHC 
energies, the addition of the extra resonances in PDG-2016+ seem to have eased the tension 
between the two flavors as suggested by the central value of PDG-2016+ that falls outside the band of PDG-2016 
and closer to unity. However, as was noted earlier with PDG-2012~\cite{Chatterjee:2013yga}, the 
discussion on flavor hierarchy though triggered by the LHC data, seems most prominent between 
$\sNN\sim10-100$ GeV. The qualitative nature of the flavor hierarchy structure remains in both the hadron 
spectra. The $T_s/T_{ns}$ shows a broad peak like structure while $V_sT_s^3/V_{ns}T_{ns}^3$ a trough in 
this range of beam energies. The central values seem to suggest a flip in the hierarchy 
for $\sNN<10$ GeV with PDG-2016+. However, again the systematic uncertainties are too large to make 
a conclusive statement on that. Such non-monotonic behaviour between $\sNN\sim10-100$ GeV hints at possible 
delayed freezeout of the non-strange sector~\cite{Chatterjee:2013yga}. It is to be noted that the beam energy 
scan data from STAR~\cite{Adamczyk:2017iwn} shows very interesting trend in mean transverse mass $\la m_T\ra$ - $m$ 
similar to $VT^3$ at these energies whose origin is yet to be understood completely~\cite{VanHove:1982vk, Mohanty:2003fy}. 
It would be interesting in the future to investigate whether these distinct trends in flavor hierarchy and 
transeverse mass at these beam energies share a common physics origin. 

Thus, we find that the systematics of the hadron spectrum mostly influence the freezeout $T$. The dimensionless 
parameters $\mu_B/T$ and $VT^3$ are comparatively more stable towards such systematics. The addition of more 
resonances mostly reduce the $T$. This happens in two ways. Firstly, with the addition of more strange resonances 
the required $\mu_S/\mu_B$ from the strangeness neutrality condition shifts to lower $T$~\cite{Bazavov:2014xya, 
Alba:2017mqu}. This cooling only occurs for the strange sector. The second way is through the feedown of these 
addtional resonances. The feeddown to the non-strange sector is from all resonances while that to the 
strange sector is only from the strange resonances. This would mean while the first factor cools only the strange 
$T$, the second factor cools both flavors, albeit more strongly the non-strange $T$. This is the reason 
behind the survival of the flavor hierarchy on addition of the extra resonances.

\begin{figure}
 \begin{center}
 \includegraphics[scale=0.42]{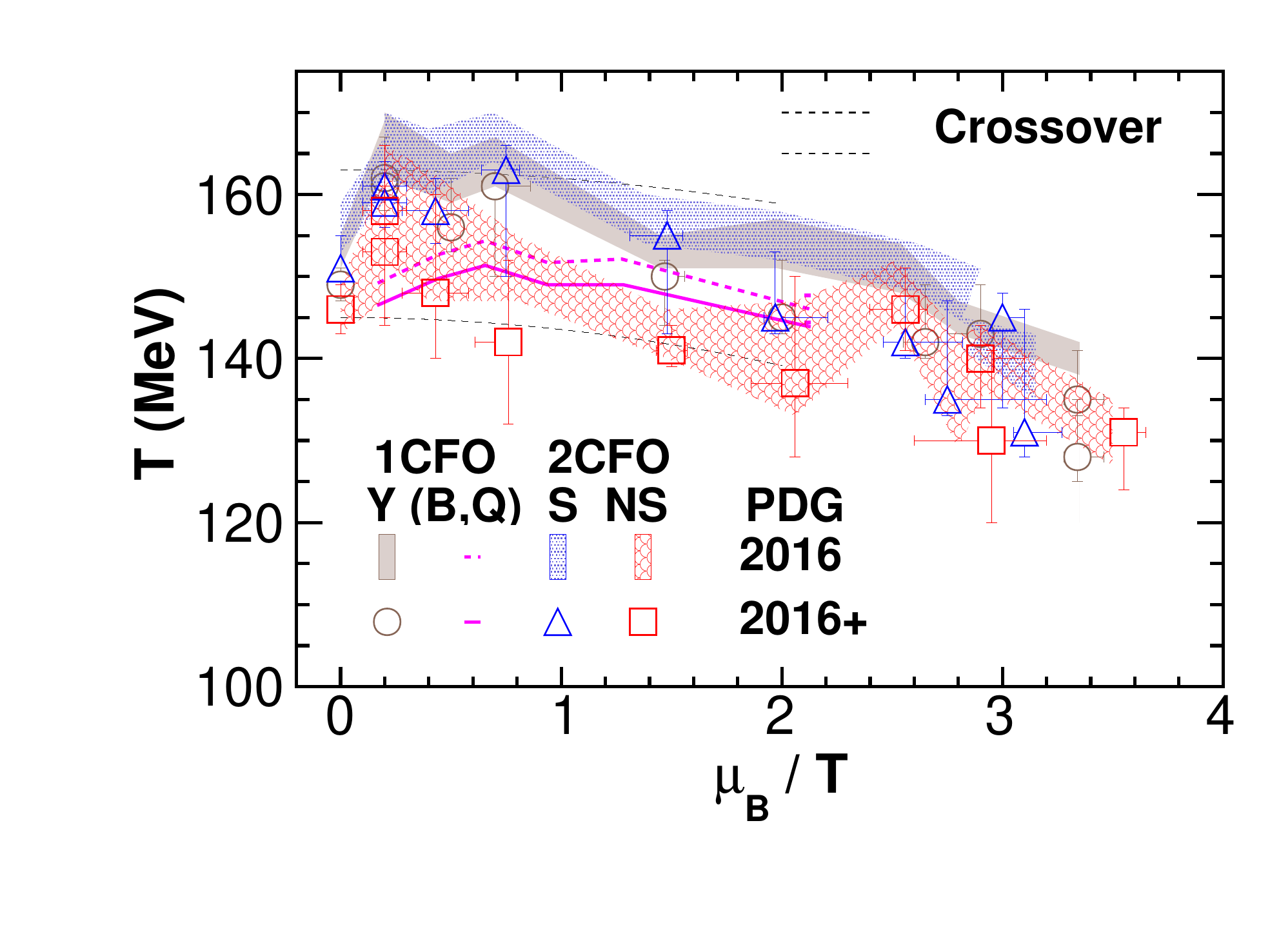}
 \caption{(Color online) $T-\mu_B/T$ plane: Freezeout coordinates in heavy ion collisions have been compared with 
 the QCD crossover region as estimated in lattice QCD computations~\cite{Bazavov:2011nk,Kaczmarek:2011zz, Endrodi:2011gv, 
 Bonati:2015bha, Bellwied:2015rza}.}
 \label{fig.tvsmubyt}
 \end{center}
\end{figure}

We have finally plotted all the freezeout $T$ and $\mu_B/T$ together in Fig.~\ref{fig.tvsmubyt}. The 
estimate of the QCD crossover transition from the hadron to the quark gluon plasma (QGP) phase from lattice 
QCD approach is also shown for comparison by the two dashed lines. The dashed lines correspond to $T=163$ MeV 
(upper) and $145$ MeV (lower) at $\mu_B=0$~\cite{Bazavov:2011nk} and has curvature $\sim0.006$ (upper)~\cite{Kaczmarek:2011zz, 
Endrodi:2011gv} and $\sim0.01$ (lower)~\cite{Bonati:2015bha, Bellwied:2015rza} that cover estimates of the 
QCD crossover region from lattice QCD calculations. The PDG-2016 spectrum yields freezeout $T<165$ MeV (central values). 
This results in a consistent picture between the haronization surface estimated on the lattice and the 
freezeout temperature extracted. The addition of the suspected 
resonances in PDG-2016+ further shifts the freezeout surface into the hadron phase. Currently, lattice approaches 
can provide results upto $\mu_B/T\sim2$. However, the freezeout parameters at lower beam energies that 
result in $\mu_B/T$ upto $\sim4$ suggest no abrupt change in curvature and probably a smooth continuation of the 
hadronization surface into higher baryon densities. It would be interesting to check how these results modify 
on using a different variant of HRG where inter-hadron interactions are considered through attractive and 
repulsive channels~\cite{Albright:2015uua,Vovchenko:2016rkn,Huovinen:2017ogf,Vovchenko:2017xad}.

\section{Summary}\label{sec.summary}
The analysis of the hadron yields within the hadron resonance gas framework provides an access to the 
freezeout surface. However such endeavours always suffer from the systematic uncertainties of the input 
hadron spectrum. The standard rule of thumb is to include all the confirmed resonances listed in the PDG. 
Studies of the hadron spectrum on the lattice as well as in quark models hint at the existence of many more 
resonances which are yet to be confirmed experimentally. Such missing resonances seem to have a considerable 
effect on thermodynamics and could also affect the phenomenology of freezeout in heavy ion collisions. In this 
work we have updated the thermal model fits with the latest PDG 2016 confirmed 
resonances lisiting as well as gauged the effect of those which are included in PDG 2016 but yet to be 
confirmed. The most significant effect is on the extracted temperature that drops 
by $\sim5\%$ while the other parameters like the phase space volume $VT^3$ and baryon fugacity $\mu_B/T$ are 
almost insensitive to such change in spectrum. These new freezeout temperatures with the updated hadron 
spectrum in both the freezeout schemes 1CFO and 2CFO are within the upper bound served by the lattice estimate 
of the hadronization temperature.

The drop in temperature is further confirmed by the analysis of the data on conserved charge fluctuation. 
The freezeout temperature extracted from the conserved charge fluctuation data is always smaller than that extracted from 
the data on mean hadron yields within 1CFO. However, the non-strange freezeout temperature extracted within 2CFO 
from yield data are consistent with that extracted from scaled variance of net $B$ and net $Q$ data. The 2CFO scenario 
can be confirmed from the fluctuation data with the availability of data on higher moments of net 
$S$ and its correlations with other flavors~\cite{Chatterjee:2016mve,Noronha-Hostler:2016rpd,Yang:2016xga,Zhou:2017jfk}. 
An obstacle in this regard is that $\Lambda$ is currently not measured on an event by event basis and one has to rely 
on the data of only charged kaons~\cite{Adamczyk:2017wsl}. Atleast at top STAR and LHC energies when the fireball 
is dominantly mesonic, we expect the data on net kaon fluctuation to be a good proxy for net $S$~\cite{Zhou:2017jfk}.

Finally, one could ask how future updates on the hadron spectrum will influence our results. A comparison of the 
thermal fits for PDG-2016 and PDG-2016+ suggests that while the dimensionless parameters $\mu_B/T$ and $VT^3$ stay 
stable to such modifications of the hadron spectrum, the freezeout $T$ is expected to fall further. The cooling 
happens pimarily due to two factors: 1. {\it The shift in $\mu_S/\mu_B$ vs $T$ plot towards lower $T$} - This shift is 
controlled by strange resonances. The newly added strange hadrons shift this plot further towards smaller $T$. In a 
1CFO scenario where there is one freezeout $T$, this cools both strange and non-strange sectors. Within 2CFO scenario, 
this cools only the strange sector and reduces flavor hierarchy. As seen from Fig.~\ref{fig.musbyb}, the PDG-2016+ 
spectrum already has a $\mu_S/\mu_B$ vs $T$ plot that is on the lower $T$ edge of the continuum lattice estimate. This 
means future hadron spectrum updates that obey this lattice estimate can not cool further through this mechanism. Since, 
PDG-2016+ already favour a flavor hierarchy in freezeout parameters as seen in Fig.~\ref{fig.sbyns}, we do not expect 
future updates on the hadron spectrum to diminish this hierarchy, and 2. {\it Feeddown from the additional resonances} - 
The feeddown from unstable additional resonances cool both non-strange and strange sectors. However, while all resonances 
feeddown to the non-strange sector, only strange resonances feed to the strange sector. Thus, this feeddown mechanism is 
expected to cool the non-strange sector more than the strange sector and hence enhance the current flavor hierarchy. 
Thus, probably our study confirms that the data prefers a hierarchial treatment of flavors. However, 2CFO where the 
hierarchy is introduced in the thermodynamic state of the fireball is not the only way to introduce flavor dependence. It 
has been shown that one could also introduce flavor dependent attractive and repulsive interactions~\cite{Alba:2016hwx}. 
It is important to construct observables to discriminate such scenarios of flavor hierarchy.

\section{Acknowledgement}
SC acknowledges discussions and collaborations on freezeout with Rohini Godbole 
and Sourendu Gupta. SC is supported by the Polish Ministry of Science and Higher 
Education (MNiSW) and the National Science Centre grant 2015/17/B/ST2/00101. BM 
acknowledges financial support from J C Bose Fellowship of DST, Govt. of India. 
SS acknowledges financial support from DAE-SRC grant, Govt. of India.

\bibliographystyle{apsrev4-1}
\bibliography{ExtraResonances}

\end{document}